\documentclass[10pt,conference]{IEEEtran}

\usepackage{cite}

\ifCLASSINFOpdf
   \usepackage[pdftex]{graphicx}
   \graphicspath{{figs/}}
   \DeclareGraphicsExtensions{.pdf,.jpeg,.png}
\else
   \usepackage[dvips]{graphicx}
   \graphicspath{{../figs/}}
   \DeclareGraphicsExtensions{.eps}
\fi
\usepackage[cmex10]{amsmath}
\usepackage{amsthm}

\usepackage{algorithmic}

\usepackage{array}

\ifCLASSOPTIONcompsoc
  \usepackage[caption=false,font=normalsize,labelfont=sf,textfont=sf]{subfig}
\else
  \usepackage[caption=false,font=footnotesize]{subfig}
\fi
\usepackage{url}

\usepackage{listings}

\begin{document}

%

\title{Ray-VR: Ray Tracing Virtual Reality in Falcor}

\newif\iffinal
\finaltrue
\newcommand{\cmtid}{6}


\iffinal

\author{\IEEEauthorblockN{Vinicius Silva}
\IEEEauthorblockA{IMPA\\
Email: dsilva.vinicius@gmail.com}
\and
\IEEEauthorblockN{Luiz Velho}
\IEEEauthorblockA{IMPA\\
Email: lvelho@impa.br}
}


%

\else
  \author{Sibgrapi paper ID: \cmtid \\ }
\fi

\maketitle

\begin{abstract}
NVidia RTX platform has been changing and extending the possibilities for real time Computer Graphics applications. It is the first time in history that retail graphics cards have full hardware support for ray tracing primitives. It still a long way to fully understand and optimize its use and this task itself is a fertile field for scientific progression. However, another path is to explore the platform as an expansion of paradigms for other problems. For example, the integration of real time Ray Tracing and Virtual Reality can result in interesting applications for visualization of Non-Euclidean Geometry and 3D Manifolds. In this paper we present Ray-VR, a novel algorithm for real time stereo ray tracing, constructed on top of Falcor, NVidia's scientific prototyping framework.
\end{abstract}



\section{Motivation}
\label{sec:intro}

Computer Graphics history has several examples of important hardware milestones. They  changed the way real time algorithms could be designed and implemented and created vast opportunities for advances in research.

One aspect of graphics cards that has been advancing consistently in the last decades is shader flexibility. In the beginning we had graphics libraries using a fixed rendering pipeline, which could only receive data and instructions from the CPU. No GPU side programming could be done at that time. This aspect was changed later, with the advent of programmable shaders. Vertex and Pixel Shaders were introduced, creating a revolution in the possibilities for real time graphics. Later on, those capabilities were increased with the exposition of more programmable rendering stages~\cite{fernando_gpu_2004}. Applications could implement Tesselation and Geometry Shaders to have access to customizable geometry resolution and primitive connectivity.

With the increment in shader flexibility, several algorithms were proposed to solve general problems using graphics hardware~\cite{pharr_gpu_2005}. The technique was to adapt the problem description to fit the rendering pipeline and the Single Instruction Multiple Data (SIMD) model that shaders use.  The next step in graphics hardware was clear: a model generalization. The result was called General Purpose Graphics Processing Units (GPGPU), a unified way to make general parallel computing using buffers in graphics memory and programming languages created specifically for that purpose~\cite{luebke_gpgpu_2006, nguyen_gpu_2007}. Examples of such languages include CUDA~\cite{nickolls_cuda_2008, sanders_cuda_2010, cook_cuda_2012}, OpenCL~\cite{munshi_opencl_2009, stone_opencl_2010, munshi_opencl_2011} and OpenACC~\cite{wienke_openacc_2012}. 

However, GPGPU applications are hard to develop in essence. Differently from shaders in the rendering pipeline, it needs explicit synchronization and memory control. This more generic model came with the costs of complexity. Nonetheless, it was explored in a vast set of problems, including but not limited to Collision Detection and Response~\cite{lauterbach_gproximity_2010, silva_lazy_2014}, Physically-based Simulation, Fluid Dynamics~\cite{riegel_fluid_2009}, Global Illumination~\cite{parker_optix_2010}, Image and Video Processing~\cite{yang_image_2008, colic_video_2010}, Segmentation~\cite{pan_medical_2008, ren_gslic_2011}, High Performance Computing and Clusters~\cite{fatica_accelerating_2009, jacobsen_mpi_2010}, Signal Processing~\cite{ujaldon_signal_2009}, Computer Vision~\cite{allusse_gpucv_2008}, Neural Networks~\cite{jang_neural_2008}, Cryptography\cite{manavski_cuda_2007}, Cryptocurrency Mining~\cite{taylor_bitcoin_2013}, Databases~\cite{bakkum_accelerating_2010}, Big Data and Data Science~\cite{chen_big_2014}. All those applications proved the model robust.

However, it was the time to migrate from hard-to-develop GPGPU to application-specific platforms for the most interesting and important problems. Machine Learning was already experiencing a revolution at that time, with exciting results coming from the association of graphics hardware and deep neural networks. The first retail graphics cards with application-specific hardware had deep neural network training and inference sections, called tensor cores. Additionally, frameworks created on top of GPGPU libraries provided a simpler API for development.

Recently, the same approach was used to create a solution for Real time Global Illumination.  The so-called RTX platform can produce faithful images using Ray Tracing (RT), which is historically known to have prohibitive performance for real time applications. This landmark creates interesting opportunities for new visualization applications. In particular, content makers for Virtual Reality (VR) can greatly benefit from the added realism to create immersive, meaningful experiences.

Thus, the demand for a VR/RT integrated solution is clear. However, realistic VR needs stereo images for parallax sensation. The obvious consequence is a duplication of the performance hit caused by ray tracing. A good algorithm should balance performance and image quality, something that can be done using RTX Ray Tracing and a proper trace policy. The recent announcement of ray tracing support for older architectures~\cite{burnes_accelerating_2019} emphasizes even more the necessity of a flexible algorithm for such task. Another point that must be taken into consideration is stereo camera registration. Depending on how the ray directions are calculated based on camera parameters, the stereo images may diverge when seen in a head mounted display (HMD).

This paper discusses the problem of integrating VR and RT, proposing a flexible solution. Section~\ref{sec:tech} describes the technological background used. Section~\ref{sec:ray_tracing_vr} contains the details of the components needed for the VR/RT integration, which is described in depth in Section~\ref{sec:integration}. Section~\ref{sec:experiments} contains the evaluation of the experiments. Finally, Section~\ref{sec:conclusion} is the conclusion.

\begin{figure*}[!t]
	\centering
	\includegraphics[width=0.8\textwidth]{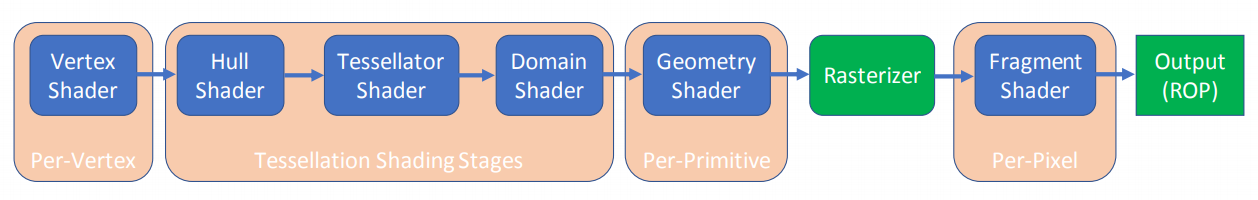}
	\caption{Rendering pipeline in depth. The blue boxes are programmable shaders and the green boxes are fixed stages. As in\cite{wyman_introduction_2018}.}
	\label{fig:raster_pipeline}
\end{figure*}

\begin{figure*}[!h]
	\centering
	\includegraphics[width=0.8\textwidth]{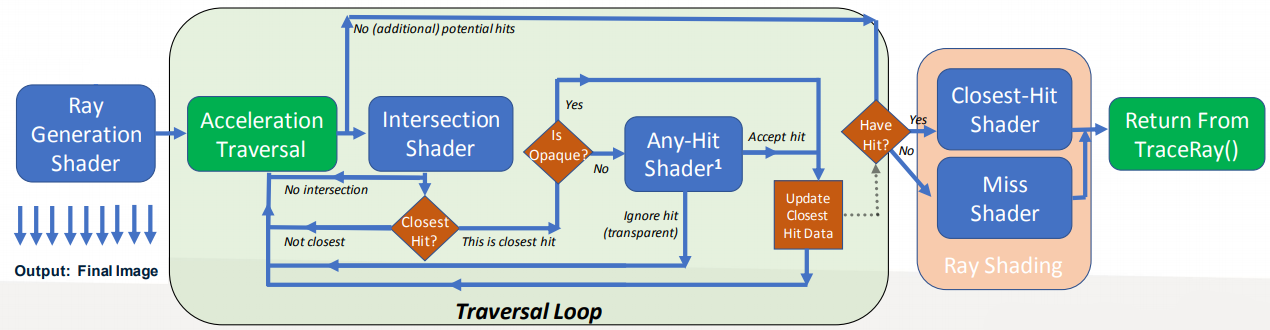}
	\caption{Ray tracing pipeline in depth. Fixed stages are in green and programmable shaders are in blue. Modified from~\cite{wyman_introduction_2018}.}
	\label{fig:ray_pipeline}
\end{figure*}

\section{Technology}
\label{sec:tech}

\subsection{RTX Ray Tracing}
\label{sec:rtx}

NVidia RTX is a hardware and software platform with support for real time ray tracing. The ray tracing code of an application using this architecture consists of CPU host code, GPU device code, the memory to transfer data between them and the Acceleration Structures for fast geometry culling when intersecting rays and scene objects.

Specifically, the CPU host code manages the memory flow between devices, sets up, controls and spawn GPU shaders and defines the Acceleration Structures. On one hand, the bottom level Acceleration Structure contains the rendering primitives (triangles for example). On the other hand, the top level Acceleration Structure is a hierarchical grouping of bottom level ones. Finally, the GPU role is to run instances of the ray tracing shaders in parallel. This is similar to the well-established rasterization rendering pipeline. 




Historically, GPUs process data in a predefined rendering pipeline, which has several  programmable and fixed processing stages. The main idea is to start with a group of stages that process the vertices, feeding a fixed Rasterizer, which in its turn generates data for pixel processing in another stage group. Finally, the result image is output by the final fixed stage. 


Currently, programmable shaders are very flexible in essence. The Vertex Shader works on the input vertices, using transformation matrices to map them to other spaces. The Hull, Tesselator and Domain Shaders subdivide geometry and add detail inside graphics memory, optimizing performance. The Geometry Shader processes primitives and mesh connectivity, possibly creating new primitives in the process. The Fragment Shader works on the pixels coming fromt the Rasterizer and the Output stage outputs the resulting image. 
Figure~\ref{fig:raster_pipeline} shows the rendering pipeline in detail.


The ray tracing GPU device code runs in a similar pipeline scheme. The differences are the stages taken. The goal of the first stages is to generate the rays. Afterwards, a fixed intersection stage calculates the intersection of the rays with the scene geometry. Then, the intersection points are reported to the group of shading stages. Notice that more rays can be created at this point, resulting in a recursion in the pipeline. The final fixed stage outputs the generated image. 


The details of the pipeline are as follows. A Ray Generation Shader is responsible for creating the rays, which are defined by their origins, directions and payloads (custom user-defined data). A call to TraceRay() launches a ray. The next stage is a fixed traversal of the Acceleration Structure, which is defined by the CPU host code beforehand. The Acceleration Traversal uses an Intersection Shader to calculate the intersections. All hits found pass by tests to verify if they are the closest hit or if they must be ignored because of transparent material. In case a transparent material is detected, the Any-Hit Shader is called for all hits so shading can be accumulated, for example. After no additional hits are found, the Closest-Hit Shader is called for the closest intersection point. In case no hits are found, the Miss Shader is called as a fallback case. It is important to note that additional rays can be launched in the Closest-Hit and Miss shaders. Figure~\ref{fig:ray_pipeline} shows an in-depth pipeline scheme. More detailed information about RTX Ray Tracing can be seen in \cite{wyman_introduction_2018} and applications can be found in \cite{haines_ray_2019}.


\subsection{Falcor}

RTX Ray Tracing can be accessed in four ways. On one hand there are the low level APIs Vulkan\cite{bailey_vulkan_2018}~\cite{sellers_vulkan_2016}, DirectX 12\cite{luna_directx_2016}~\cite{wyman_introduction_2018} and OptiX\cite{parker_optix_2010}. They provide more flexibility but less productivity. On the other hand there is {Falcor}\cite{benty_falcor_2018}, an open-source real-time rendering framework designed specifically for rapid prototyping. It has support for ray tracing shaders and is the recommended way to use RTX Ray Tracing in a scientific environment.

Falcor code and installation instructions can be found at {https://github.com/NVIDIAGameWorks/Falcor}~\cite{benty_falcor_2018}. The bundle comes with a Visual Studio solution, structured in two main components: a library project called Falcor with high level components and Sample projects which use those components to perform computations, effects or to provide tools for other supportive purposes.

Each Sample project consists at least of a main class inheriting from Renderer and a Data folder. The Renderer class defines several relevant callbacks which can be overridden as necessary. Examples include onLoad(), onFrameRender(), onGuiRender(), onMouseEvent() and so forth. The Data folder is where non-C++ files necessary for the Sample (usually HLSL Shaders) should be placed. Falcor automatically copies them at compilation time to the binary's folder so programs have no access problems.

\section{Ray Tracing VR}
\label{sec:ray_tracing_vr}

Our goal is to build a stereo-and-ray-tracing-capable renderer for VR. 
For this purpose, we will exploit the functionalities of Falcor that provide support for Stereo Rendering, Simple Ray Tracing and Global Illumination Path Tracing.

Falcor is designed to abstract scene and Acceleration Structure setup so our focus will be on describing Shader code and the CPU host code to set it up. The next subsections explain the logic for three Falcor Samples with the objective of using their components later on as building blocks for our new renderer. We refer to code in the Falcor Samples, so it is advisable to access it in conjunction with this section for a better understanding.

\subsection{Simple Ray Tracer}

HelloDXR is a simple ray tracer with support for mirror reflections and shadows only. As would be expected, the Sample specifies two ray types: primary and shadow. 

The ray generation shader \lstinline{rayGen()} is responsible of converting pixel coordinates to ray directions. This is done by a transformation to normalized device coordinates followed by another transformation using the inverse view matrix and the camera field of view. The function TraceRay() is used to launch the rays. The ray type index and the payload are provided as parameters.

The Closest-Hit Shader \lstinline{primaryClosestHit()} calculates the final color of the pixel. It has two components: an indirect reflection color and a direct color. The reflection color is calculated by \lstinline{getReflectionColor()}, which reflects the ray direction using the surface normal and shoots an additional ray in that direction. The payload has a ray depth value used to limit recursion. The direct color is the sum of the contributions of each light source at the pixel, conditioned by the shadow check  \lstinline{checkLightHit()}. If the light source is not occluded, the contribution is calculated by \lstinline{evalMaterial()}, a Falcor built-in function to shade pixels based on materials.

\subsection{PathTracing}
\label{sec:path_tracing}

The PathTracer Sample implements the Path Tracing algorithm~\cite{kajiya_equation_1986}. It uses a RenderGraph to chain four rendering steps: G-Buffer rasterization, global illumination, accumulation and tone mapping. The graph is defined in CPU host code, using \lstinline{addPass()} and \lstinline{addEdge()} to create the passes and links their input and output respectively.

Each pass has its own Shaders. The G-Buffer pass uses the rasterization pipeline to output shading data to a set of textures. More specifically, the built-in function \lstinline{prepareShadingData()} is used in a Pixel Shader to fetch and sample the material, whose data is output to the G-Buffers. Falcor default fallback Vertex Shader is used in this step.

The global illumination pass calculates direct and indirect contributions as well as shadows. It uses a ray generation shader and two ray types: indirect and shadow. Each type is defined in host code by a hit group (a closest-hit and an any-hit shader) and a miss shader. The setup is done using a descriptor and calling \lstinline{addHitGroup()} and \lstinline{addMiss()} respectively. Method \lstinline{setRayGen()} is called to define the ray generation shader as well. Each ray type must have a unique index, which is referenced in the shaders. Shadow rays have index 0 and indirect ones have index 1.

The ray generation shader controls the path tracing. Briefly, the coordinates of each pixel are used to compute random number generator seeds, which are used to calculate random directions for indirect rays. The indirect sample is chosen randomly from the diffuse hemisphere or specular direction. An analogous idea is used for direct lighting as well, which is computed by randomly choosing a light source to check for visibility. As shown in~\cite{kajiya_equation_1986}, this integration converges to the complete evaluation of the illumination of the scene.

Now, the Hit Group and Miss Shaders are described. The shadow ray Shaders are very simple. The Miss Shader sets the ray visibility factor to 1 from the default 0, which means that the ray origin is visible to the light and should be lit by it. The Any-Hit shader just checks if the ray origin has a transparent material using the built-in evalRtAlphaTest() function. Case it is a transparent material, the hit is ignored so the ray can continue its path. There is no Closest-Hit Shader for shadows since the visibility factor should change only if no objects are hit.

Analogously, indirect rays have their own Shaders. The main difference between the ray types is the Closest-Hit Shader, which calculates the direct light at the intersection point and shoot an additional ray case the depth is bellow the global threshold. The Miss Shader samples a color from the environment map, indexed by the ray direction, and the Any-Hit Shader is equal to the shadow ray's. 

The accumulation pass is also very simple. Its CPU code maintains a texture with the previous frame and ensures that accumulation is done only when the camera is static. This image is accumulated with the current frame, coming from the global illumination pass. The accumulation consists of an incremental average.

Finally, Falcor built-in tone mapping is used to adjust the colors of the image. Class \lstinline{ToneMapping} abstracts this pass.

\subsection{StereoRendering}
\label{sec:stereo}

The StereoRendering Sample is an application to render stereo image pairs using rasterization. The CPU host code ensures connection with the HMD (\lstinline{initVR()}), issues the Shaders to generate the images and transfers them to the device (\lstinline{submitStereo()}). Specifically, it maintains a struct containing the camera matrices and properties for both eyes. The geometry is drawn once, but it is duplicated inside the GPU by the Shaders. A frame buffer array with two elements is maintained for that purpose (\lstinline{mVrFbo}). When a frame finishes, each array slice has the view of one eye.

The GPU code consists of a Vertex Shader, a Geometry Shader and a Pixel Shader. The Vertex Shader (\lstinline{StereoRendering.vs.hlsl}) just passes ahead the vertex positions in world coordinates and additional rendering info such as normals, colors, bitangent, texture and light map coordinates, if available.

The projection is left to the Geometry Shader (\lstinline{StereoRendering.gs.hlsl}), which is also responsible for duplicating the geometry. It receives as input three vertices of a triangle and outputs six vertices. Each input vertex is projected twice, once for each of the view-projection matrices available at the camera struct. The geometry for each eye is output into the related array slice by setting a render target index at struct \lstinline{GeometryOut}.

Finally, the Pixel Shader (\lstinline{StereoRendering.ps.hlsl}) is very simple. It just samples material data using the built-in function \lstinline{prepareShadingData()} and accumulates the contributions of each light source using the built-in function \lstinline{evalMaterial()}.

\section{Integration}
\label{sec:integration}

Our goal is to develop a new renderer that combines the capabilities described in the previous section. It should be capable of stereo rendering and ray tracing in real time. This section describes the process and the possible choices and alternatives to address the problems encountered.

\subsection{Stereo Convergence}
\label{sec:stereo_convergence}

One key problem of integrating VR and RT is the stereo image registration. Depending on how this process is done, the images may diverge and it can be impossible for the human vision to focus correctly on the scene objects. This phenomenon may result in viewer discomfort or sickness.

To understand the ray generation process it is good to think about perspective projection and the several related spaces it involves. \cite{pharr_physically_2016} contains an exceptional explanation of this topic, which will be summarized here.

\begin{figure}[!h]
	\centering
	\includegraphics[width=.4\textwidth]{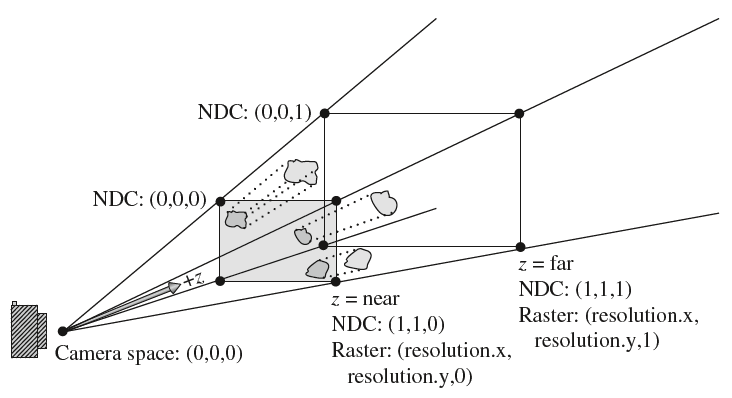}
	\caption{Several camera-related coordinate spaces: camera, normalized device coordinates and raster. As in~\cite{pharr_physically_2016}.}
	\label{fig:spaces}
\end{figure}

Conceptually, the process consists of a chain of transformations starting at the world space, passing through the camera space and the normalized device coordinate space and ending at the raster space. The camera space is the world space with a translated origin to the camera position. The normalized device coordinate space is the camera space with the near and far planes transformed. The near plane is at the square with top-left corner at (0,0,0) and bottom-right corner at (1,1,0) and the far plane is at the square with top-left corner at (0,0,1) and bottom-right corner at (1,1,1). Finally, the raster space is the normalized device coordinate space scaled by the image resolution. Figure~\ref{fig:spaces} shows how the spaces relate to each other.

The transformation for a projection camera can be constructed in two steps. First, building a canonical perspective matrix with distance to the near plane $n$ and distance to the far plane $f$. The projected coordinates $x'$ and $y'$ are equal to the original ones divided by the $z$ coordinate. $z'$ is remapped so the values in the near plane have $z' = 0$ and the values in the far plane have $z' = 1$:

\[
\begin{gathered}
	x' = x/z \\
	y' = y/z \\
	z' = \frac{f(z-n)}{z(f-n)}.
\end{gathered}
\]

This operation can also be encoded as a $4x4$ matrix using homogeneous coordinates:

\[
	\begin{bmatrix}
    	1 & 0 & 0 & 0 \\
	    0 & 1 & 0 & 0 \\
    	0 & 0 & \frac{f}{f - n} & \frac{-f*n}{(f - n)} \\
	    0 & 0 & 1 & 0
	\end{bmatrix}
\]

As a side note, the original position of the projection planes would be important for rasterization because the map of $z$ to $z'$ is not linear, what could possibly result in numerical issues at depth test, for example. However, we are only interested in the projection directions for ray tracing, thus those distances can be totally arbitrary.

\begin{figure}[!h]
	\centering
	\includegraphics[width=.3\textwidth]{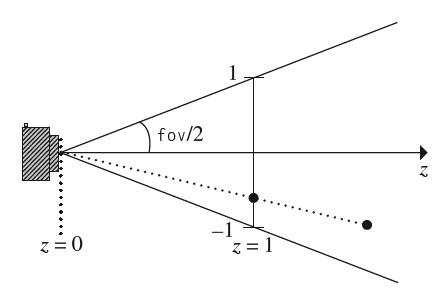}
	\caption{Relation between a field of view and normalized device coordinates. As in~\cite{pharr_physically_2016}.}
	\label{fig:perspective_projection}
\end{figure}

The second step is scaling the matrix so points inside the field of view project map to coordinates between $[-1,1]$ on the view plane. For square images, both $x$ and $y$ lie between the expected interval after projection. Otherwise, the direction in which the image is narrower maps correctly, and the wider direction maps to a proportionally larger range of screen space values. The scaling factor that maps the wider direction to the range $[-1,1]$ can be computed using the tangent of half of the field of view angle. More precisely, it is equal to:

\[
	\frac{1}{tan(\frac{fov}{2})},
\]
as can be seen in Figure~\ref{fig:perspective_projection}.

To launch rays from pixels we use the inverse transformation chain. We start at the raster space, passing through the normalized device coordinate and camera spaces and ending at the world space. More specifically, to compute a ray direction we must convert the raster coordinates of its associated pixel to normalized device coordinates, scale by the reciprocal of the factor used to map the field of view to the range $[-1,1]$ and use the inverse view transformation matrix to map the result to the world space. The conversion from raster coordinates $r \in [d_x,d_y]$ to normalized device coordinates $n \in [-1,1]$, given the image dimensions $d = (d_x,d_y)$, is expressed by the following equation:

\[
n = \frac{2(r + 0.5)}{d} - 1;
\]

which is composed of a normalization by the image dimensions and operations to map the resulting image space from the interval $[0,1]$ to $[-1,1]$.

The remains of the transformation chain can be done in a optimized way, using a precomputed tangent of half the field of view angle $f$ (in dimension $y$), the aspect ratio $a$ and the basis vectors of the inverse view matrix $I = [ u | v | w ]$. The operation is done by the expression:

\[
\begin{gathered}
	normalize(a f (n_x u) - (f (n_y v)) - w),
\end{gathered}
\]

As can be seen, this expression transforms the normalized device coordinates using the parts of the inverse view matrix that would affect each of the coordinates and scales them using the tangent of the field of view for dimension $y$. The scale value is corrected by the aspect ratio for the $x$ dimension.

In our tests we could not generate correctly registered stereo images using this optimized expression, because it does not take into account stereo rendering. For  this reason, we used two other approaches: the inverse of the projection matrix, and a rasterization G-Buffer pre-pass. Both options ensure correct stereo images, with different pros and cons. The first option does not need any additional rasterization pass or memory for the required texture. However, the G-Buffer provides more flexibility for the algorithm as will be discussed in Section~\ref{sec:ray_tracing_overhead}. It is important to note that the positions in the texture are equivalent to intersection points of rays launched from the camera position. This property comes from the fact that the camera position is equivalent to the projection center and each ray is equivalent to the projection line for the associated pixel.

\subsection{Ray Tracing Overhead}
\label{sec:ray_tracing_overhead}

The major drawback of usual stereo rendering is the overhead caused by the additional camera image. This problem is emphasized even more in ray tracing, which demands heavy computation to generate the images. Several techniques have been proposed to address this issue. They usually create the additional image by transforming the contents of the original or by temporal coherence using previous frames. However, artifacts not present when the scene is rendered twice can be introduced in the process.

The RTX platform opens new ways to explore this problem. Additionally, the extension of the ray tracing support for older graphics card architectures~\cite{burnes_accelerating_2019} encourages new algorithms based on smart ray usage. We benefit from Falcor's design to explore and evaluate the possibilities using a methodology based on fast cycles of research, prototyping, integration and evaluation. The result is a list of several possible approaches, generated by changing component routines of a ray tracing algorithm. In summary, those changes result from the following questions.

\begin{enumerate}
	\item How the first intersections are calculated?
		\begin{enumerate}
			\item Rays shot from camera position.
			\item G-Buffer pre-pass.
		\end{enumerate}
	\item Which effects are applied?
		\begin{enumerate}
			\item Direct light and shadows only.
			\item Perfect-mirror specular reflections and shadows only.
			\item Path tracing.
		\end{enumerate}
\end{enumerate}

The different algorithms are created by combining the different functionalities described in Section~\ref{sec:ray_tracing_vr}.
We start by integrating Simple Ray Tracing and Stereo Rendering.
On one hand, Stereo Rendering includes all logic to connect with the HMD and to control the data flow between the ray tracing shaders and the device. 
On the other hand, Simple Ray Tracing features a ray tracing shader, which is modified to launch rays based on two view matrices or two position G-Buffers, one for each eye. On the G-Buffer case a rasterization pre-pass is also performed, analogously to the Path Tracing.

Next, PathTracer's components are integrated, enabling better control of the effects applied. We benefit from Falcor's RenderGraph, which is extremely useful for algorithms with multiple rendering passes. The changes needed are listed next.

\begin{enumerate}
	\item Adding an additional mirror ray type, equivalent to the primary ray type.
	\item Including a function to compute direct light with shadows only. If the G-Buffer is available, the direct contribution comes for free from it.
	\item Adding a branch in the Ray Generation Shader to choose between the effects: raster, direct light plus shadows, specular reflections and path tracing.
\end{enumerate}

An interesting question arises when we analyse the current algorithm. A ray tracing procedure with a G-Buffer pre-pass is actually a hybrid algorithm based on both rasterization and ray tracing. What if we extrapolate this hybrid paradigm and allow materials to be raster or ray-traced in the scene? This question generated an additional change in the integrated renderer. We introduced a material ID to enable support for per-material effect selection. With this feature, an user can control performance by changing the material IDs of objects in the scene from more complex to simpler effects and vice versa. The final algorithm is very flexible and suited for stereo ray tracing or for older graphics card architectures, environments where performance matters.

\section{Evaluation}
\label{sec:experiments}

In Section~\ref{sec:integration} we pointed out component routines that could be changed in a ray tracing algorithm to achieve different levels of performance and image quality. In this section we are interested in measure and evaluate those changes. Our goal is to find the best solution of compromise between those two indices.

The evaluation methodology consists of several tests using three
scenes: Falcor Arcade~\cite{benty_falcor_2018}, Epic Games Unreal
Engine 4 Sun Temple \cite{epic_orca_2017} and Amazon Lumberyard Bistro
\cite{amazon_orca_2017}. The Arcade is a very simple scene with
minimal geometry, used as a toy example. The Temple and the Bistro are
part of the {Open Research Content Archive  (ORCA)} 
and are more dense, with sizes comparable to scenes actually encountered in games and VR experiences.

All tests are done in a PC with an Intel(R) Core(TM) i7-8700 CPU, 16GB RAM, a RTX 2080 and a HTC Vive. The target performance is 90 fps, the HMD update frequency. Depending on how far the application is from this value, the device starts to reproject images and to lose frames, what can result in motion sickness for the user. Everything that could perform above 90 fps is also capped to that value. In our tests, users reported that performances near 45 fps are good, with minimal hickups noticed due to reprojection. Timings below this value started to feel uncomfortable. The values include the image transfer to the HMD.

\subsection{G-Buffers and Stereo}
\label{sec:stereo_test}

The first test evaluates the impact of the G-Buffer pre-pass and stereo rendering. As discussed in Section~\ref{sec:integration}, the hybrid algorithm depends on G-Buffers, so it is important to assess their viability early on. The methodology is:

\begin{enumerate}
    \item We have two controls in the experiment: a mono raster shader and a stereo raster shader.
    \item All ray tracing materials use the mirror-like reflection shader with one bounce.
    \item When stereo is not enabled, only the images for the left eye are generated.
\end{enumerate}

Tables \ref{tab:test1_temple} and \ref{tab:test1_bistro} shows the results for the Temple and the Bistro respectively. The Arcade runs at 90 fps in any case because of its simplicity.


\begin{table}[h!]
\centering
    \begin{tabular}{ |c|c|c| } 
        \hline
        Algorithm & Stereo & FPS \\
        \hline
        
        Raster & No & 90 \\ 
        Raster & Yes & 90 \\
        Inverse matrix & No & 90 \\ 
        Inverse matrix & Yes & 45 \\ 
        G-Buffer & No & 90 \\
        G-Buffer & Yes & 45 \\
        \hline
    \end{tabular}
    \caption{G-Buffer pre-pass and stereo imaging impact for Temple.}
    \label{tab:test1_temple}
\end{table}

\begin{table}[h!]
\centering
    \begin{tabular}{ |c|c|c| } 
        \hline
        Algorithm & Stereo & FPS \\
        \hline
        
        Raster & No & 80 \\ 
        Raster & Yes & 45 \\
        Inverse matrix & No & 80 \\ 
        Inverse matrix & Yes & 45 \\ 
        G-Buffer & No & 80 \\
        G-Buffer & Yes & 45 \\
        \hline
    \end{tabular}
    \caption{G-Buffer pre-pass and stereo imaging impact for Bistro.}
    \label{tab:test1_bistro}
\end{table}

As expected, the stereo rendering is the bottleneck and the G-Buffer overhead is negligible in comparison with it. Thus, our proposal is to use the hybrid algorithm to customize scenes and balance the indices. The next step is to measure how materials interact with them.

\subsection{Materials}
\label{sec:materials}

The second test focuses on material effects. The idea is to use the hybrid algorithm and change the material ids on-the-fly to balance quality and performance. The methodology is:

\begin{enumerate}
    \item Stereo is always enabled.
    \item The camera position is fixed.
    \item Materials are changed to balance image quality and performance. There are three possibilities.
        \begin{enumerate}
            \item Raster.
            \item Raster and ray-traced shadows.
            \item Ray-traced mirror-like reflections and shadows.
        \end{enumerate}
\end{enumerate}

The application supports an additional Path Tracing shader. However, as discussed in Section~\ref{sec:path_tracing}, the algorithm needs to accumulate samples over frames from a static camera to eliminate noise. This restriction is hard to be imposed in a VR experience, where the camera is controlled by the user's head movement. Thus, we will not be using this shader in the experiments.

\begin{figure}[!thb]
	\centering
	\includegraphics[width=\linewidth]{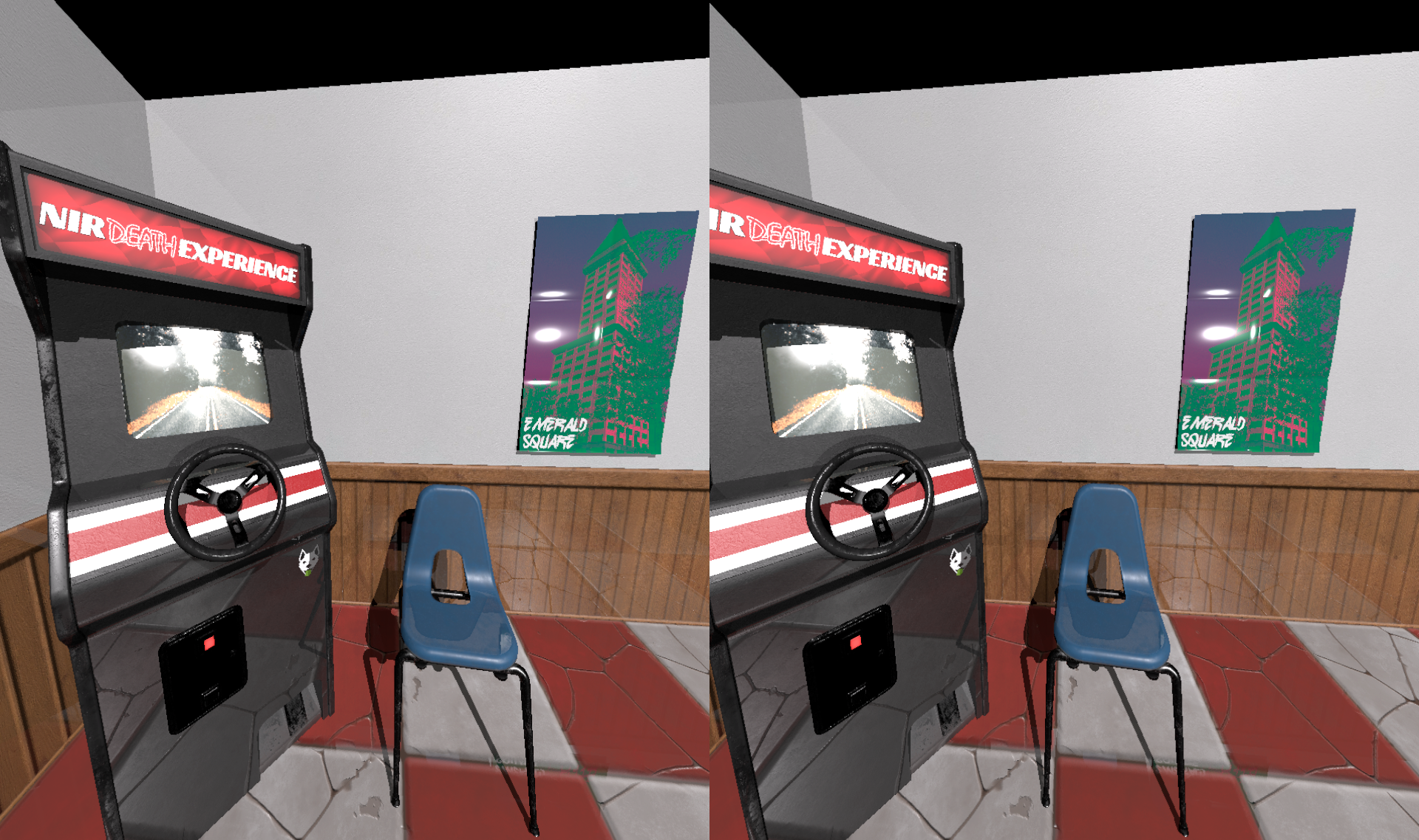}
	\caption{Arcade: ray-traced reflections and shadows.}
	\label{fig:arcade_near_mirror}
\end{figure}

\begin{figure}[!thb]
	\centering
	\includegraphics[width=\linewidth]{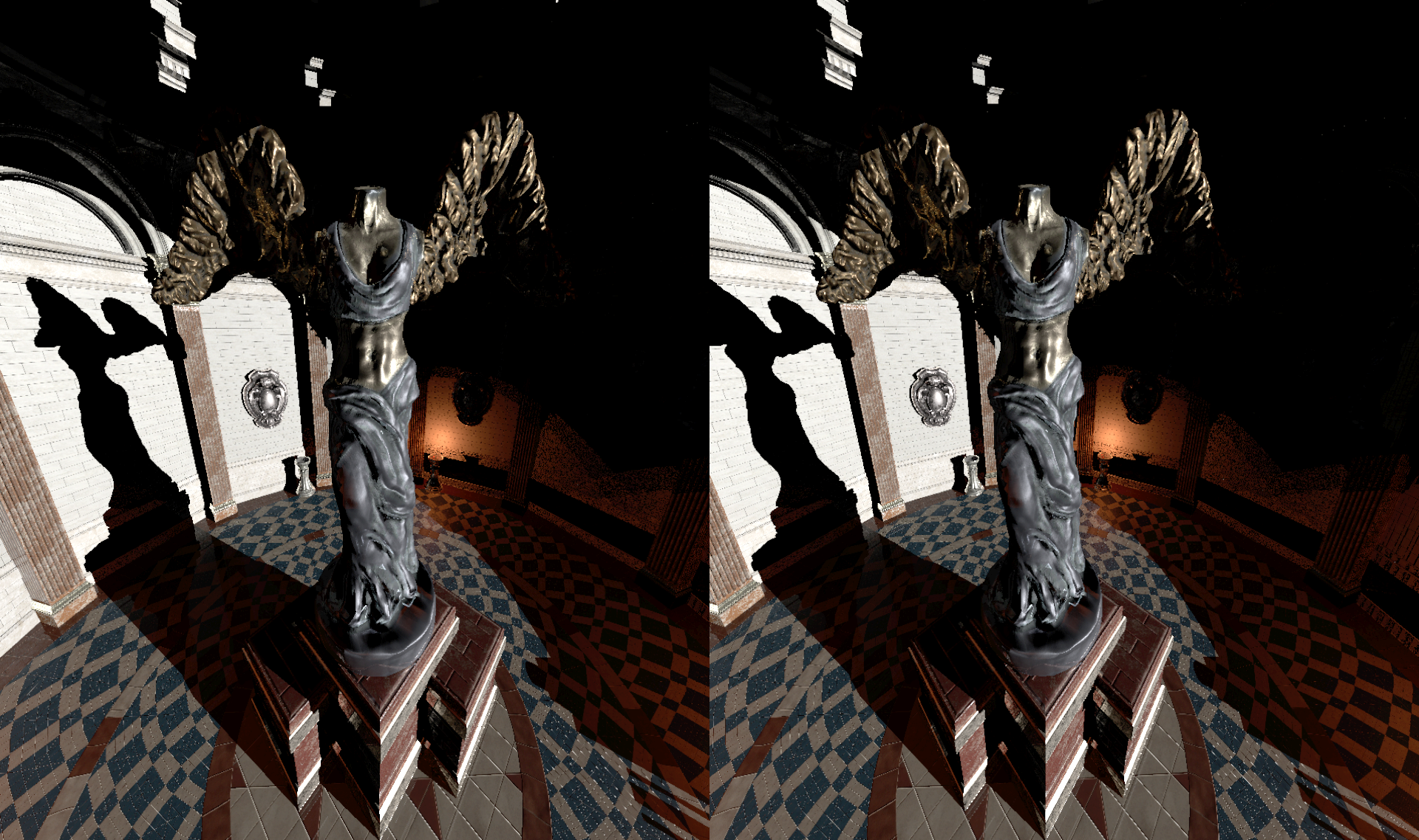}
	\caption{Sun Temple: ray-traced mirror-like reflections and shadows on everything.}
	\label{fig:temple_near_statue_all_mirror}
\end{figure}

\begin{figure}[!thb]
	\centering
	\includegraphics[width=\linewidth]{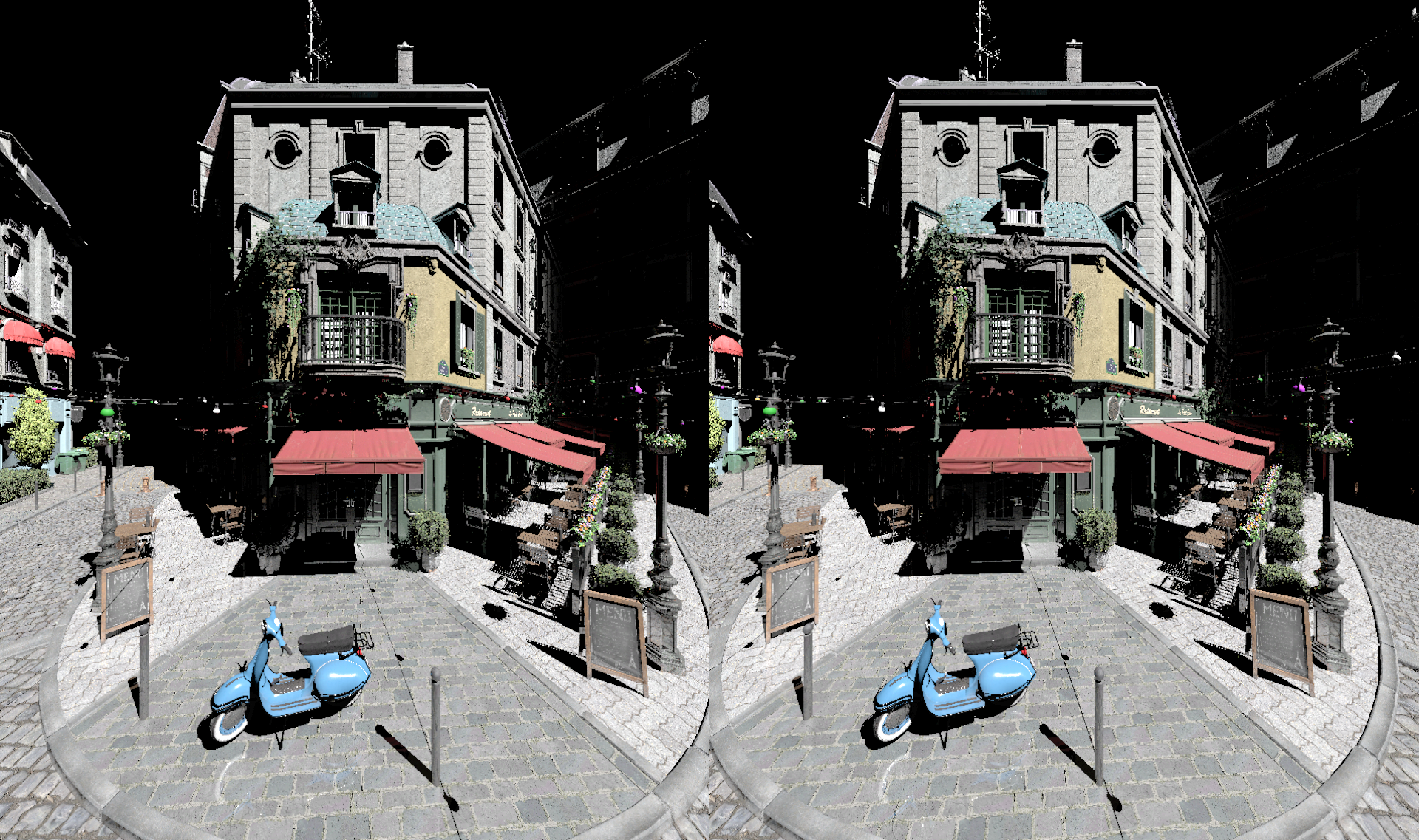}
	\caption{Bistro: ray-traced mirror-like reflections and shadows.}
	\label{fig:bistro_mirror}
\end{figure}

The Arcade stays at 90 fps regardless of effect choice so we only show the best result in Figure~\ref{fig:arcade_near_mirror}, obtained when using the ray tracing shader with shadows and reflections. In Figure~\ref{fig:temple_near_statue_all_mirror} we show the results for the Sun Temple, while Figure~\ref{fig:bistro_mirror} contain the results for the Bistro. Moreover, Tables~\ref{tab:temple} and \ref{tab:bistro} quantify the performance for each case.






\begin{table}[h!]
\centering
    \begin{tabular}{ |c|c| } 
        \hline
        Effect & FPS \\
        \hline
        
        Raster & 90 \\ 
        Raster + ray-traced shadows & 90 \\
        Ray-traced reflections on statues and wall decorations & 75 \\ 
        Ray-traced Reflections on everything & 45 \\
        \hline
    \end{tabular}
    \caption{Effect impact for the Sun Temple.}
    \label{tab:temple}
\end{table}

\begin{table}[h!]
\centering
    \begin{tabular}{ |c|c| } 
        \hline
        Effect & FPS \\
        \hline
        
        Raster & 45 \\ 
        Raster + ray-traced shadows & 45 \\
        Ray-traced Reflections and shadows & 45 \\
        \hline
    \end{tabular}
    \caption{Effect impact for the Bistro.}
    \label{tab:bistro}
\end{table}

\pagebreak

The use of effects drastically changes the mood and fidelity of the
scenes, resulting in a much better immersive experience. It is
important to remark that Tables~\ref{tab:temple} and \ref{tab:bistro}
only show a small subset of all possible material setups. Those are
the ones with minimal interaction to change the ids. However, with
proper material id tweak, the application can achieve even faster
frame rates while maintaining image quality. We could think an
automatic algorithm to set ids based on importance values given by
artists and distance of objects from the camera. Non-important or very
distant objects could be set as raster or raster plus shadows, instead
of other heavier materials.

\section{Conclusion}
\label{sec:conclusion}

In this paper we presented Ray-VR, a very flexible algorithm to integrate real time RT in VR.  At the time of manuscript submission, even Epic Games and Unity Technologies, big players in the VR market, do not have support for real time VR/RT in their game engine solutions. As far as we know, Ray-VR is the first algorithm to be successful at such task.

Ray-VR performance is very flexible in essence. It can adapt a VR experience to different hardware constraints. High performance devices can benefit from high quality ray-traced images, creating more immersive environments. However, other devices can still run the experience in real time, but with less effects.

The algorithm is also totally compatible with current VR creation workflow. The user interaction needed to change material ids is straight-forward, suited for artists at asset creation time, for developers at development time and for designers at testing time.

The human interaction needed to change the material ids can also be considered a limitation, however. Ideally, we want an algorithm that changes the material ids automatically. With this in mind, we briefly described improvements that could converge to a solution in Section~\ref{sec:materials}. The artists could assign importance values to the assets at creation time. This methodology in conjunction with other heuristics such as object distance, for example, could result in an automatic algorithm for material id setup. It could also optimize the importance value based on the original ones given by the artists and a given fps budget. Ray-VR is flexible enough to support such operations after small changes in the current algorithm.

An intuitive example is the statue at the Sun Temple. It is by far the most important asset in the scene and could have a high importance value. Walls for example, could receive much less attention, since they usually are part of the background of the scene. A more sophisticate attempt could be to create a neural network to learn how to set the material ids with scene examples in order to optimize performance and image quality.




\IEEEpeerreviewmaketitle




\bibliographystyle{IEEEtran}
\bibliography{ray_vr}

\end{document}